# Light-driven permanent transition from insulator to conductor in $Ga_2O_3$

F. A. Selim[1,2,*], D. Rana,[1] S. Agarwal,[1,2] M. Islam,[1,2] A. Banerjee,[3] B. P. Uberuaga,[3] P. Saadatkia,[1,2] P. Dulal,[1] N. Adhikari,[1] M. Butterling,[4] M. O. Liedke,[4] A. Wagner [4]

[1] Department of Physics and Astronomy, Bowling Green State University, Bowling Green, Ohio, USA 43403.

[2] Center for Photochemical Sciences, Bowling Green State University, Bowling Green, Ohio, USA 43403.

[3] Materials Science and Technology Division, Los Alamos National Laboratory, Los Alamos, NM 87545, USA.

[4] Institute of Radiation Physics, Helmholtz-Zentrum Dresden-Rossendorf, Bautzner Landstr. 400, 01328 Dresden, Germany.

## Abstract

The transition from insulator to conductor can be achieved in some materials but requires modification of both the arrangement of atoms and their electronic configurations. This is often achieved by doping. Here we reveal a mechanism the lattice may adopt to induce such a transition. We show that limited exposure to sub-bandgap light caused a permanent transition from an insulator state to a conductor state in the insulating oxide $Ga_2O_3$ with 9-orders of magnitude increase in electronic conduction. Photoexcitation modifies the charge state of an O-vacancy and the redistribution of the localized electrons, leading to a massive structural distortion in the lattice that is shown to be the underlying mechanism. It modifies density of states and introduces new stable states with shallower energy levels, leading to this intriguing behavior. We suggest that this mechanism may occur in other wide bandgap metal oxides leading to drastic modification in their electronic properties.

Corresponding should be addressed to:

[*] **faselim@bgsu.edu**

**Introduction**

When light is impinged on a semiconductor material, charge carriers – electrons and holes – may be generated resulting in an enhancement in conductivity. If the energy of the incident photons is greater than the band gap of a semiconductor, it excites an electron from the valance band to the conduction band, a phenomenon called intrinsic photoconductivity. On the other hand, if the energy of the photon is less than the band gap, it may excite electrons from defect levels to the conduction band enhancing conductivity. This is referred as extrinsic photoconductivity[1]. In either case, if the conductivity persists after turning off the photo excitation, it is known as persistent photoconductivity. In this situation, when the electron-hole pairs are generated, there must be microscopic or macroscopic potential barriers that separate charge carriers and reduce the probability of recombination between them, resulting in enhanced conductivity for a longer period of time[2].

Persistent photoconductivity at room temperature has been primarily reported in hetero-structures of semiconductors and in a few bulk materials[2, 3]. In this work we report an extrinsic persistent photoconductivity behavior in bulk $Ga_2O_3$ and a surprising permanent transition from the insulator state to the conductor state upon exposure to sub-band gap light for a limited period of time. First, $Ga_2O_3$ bulk crystals exhibited massive persistent photoconductivity upon exposure to sub-bandgap light of much lower energy than the band gap. Then, the induced meta-stable states created by light became stable simply by increasing the photoexcitation time, leading to a permanent transition from the insulator state to the conductor state. To our knowledge, such behavior has never been reported in any material system and could have vast implications on the material properties and applications.

Gallium oxide ($Ga_2O_3$) is the widest band gap transparent (up to UV-C range) semiconducting oxide known so far[4-25]. Its ultra-wide band gap (~ 4.5- 4.9 eV)[6] may lead to unusual electronic phenomena. Due to this wide band gap, UV-C transparency, and excellent thermal and chemical stability, it has numerous potential applications in power and high voltage devices, Schottky diodes, field effect transistors, gas sensors, phosphors and electroluminescent devices, UV photo detectors and more[7-11, 24]. $Ga_2O_3$ exhibits polymorphism, denoted by α, β, γ, δ, and ε[12], with β-$Ga_2O_3$ being the most stable phase from room temperature to its melting point[13]. As the most stable form, β-$Ga_2O_3$ is also the most studied polymorph. It crystallizes into a monoclinic structure with space group C2/m and lattice parameters a = 12.2140 Å, b = 3.03719 Å, c = 5.7819 Å and β= 103.83° [13]. It contains both octahedral and tetrahedral cation sites in equal numbers. As $Ga_2O_3$ has a wide band gap, it is an insulator at room temperature, but electron conduction has been reported when synthesized under reducing condition[5]. This conductivity is often attributed to oxygen vacancies, though recent theoretical calculations showed that oxygen vacancies are deep states and cannot provide conduction electrons[26]. It has been also proposed that silicon, which is a major impurity in $Ga_2O_3$, might be the cause of electron conductivity[12]. The effective hole condition in $Ga_2O_3$ has not been reported; theoretical calculations show that the valance band is flat, indicating larger effective mass for holes, making p-type conductivity difficult[6].

**Results and Discussion**

$Ga_2O_3$ single crystals were illuminated by sub-bandgap light and the conductivity and carrier density were measured during illumination. Figure 1 illustrates the dependence of photoconductivity in undoped β-$Ga_2O_3$ single crystals on photoexcitation energy and intensity. In Fig. 1b, the photoconductivity is plotted versus light intensity for photon energies of 3.39, 3.22, 3.1, 2.69 eV, revealing that the conductivity abruptly increases at low photon intensity and quickly saturates. In Fig. 1c, the photoconductivity and photo-induced charge carrier density are plotted as a function of photon energy. Before each measurement, the sample was heated at 400 °C for 1 hour to retain its initial dark conductivity, and excitation was carried at the same photon intensity for all energies. It is interesting to note that all sub-band gap photo-excitations led to an increase in conductivity, even at low energies of 1.9 and 1.45 eV. The maximum photoconductivity occurs at 3.1 eV, which is much smaller than 4.5/4.9 eV, the band gap energy of $Ga_2O_3$[6]. The increase of conductivity with sub-band gap photoexcitation can be explained due to the excitation of electrons from a localized state in the gap to the conduction band as shown in Fig. 1a. In contrast to undoped $Ga_2O_3$, Fe- and Mg-doped samples behave differently when exposed to light. Both Fe and Mg-doped crystals showed decreased conductivity when exposed to 400 nm and 365 nm light (Fig. 2). This indicates that common impurities in $Ga_2O_3$ such as Fe is not behind this photoconductivity.

After tuning off the photo-excitation, the sample illuminated by 3.1 eV shows significant persistent photoconductivity. To calculate the associated potential barrier that prevents the recapture of charge carriers by their centers after light is turned off and is thus the ultimate origin of the persistent photoconductivity[3,27], the photoconductive sample was annealed at various temperatures (from 300°C to 390°C) for 10 minutes at each temperature inside the Hall-effect chamber. After each anneal, the sample was then cooled to room temperature and the electrical conductivity and carrier density were measured. The steps of the experimental procedure are illustrated in Fig. 3a. Figure 3b show how the conductivity and charge carrier density of the sample vary with annealing temperature and Figure 3c is a corresponding graph for the natural logarithm of charge carrier density vs reciprocal thermal energy (1/kT). Using the Arrhenius equation, $n = Ae^{-E_{th}/kT}$, the slope of the best-fit line gives a thermal energy barrier of about $0.157 \pm 0.04$ eV. In Fig. 3d, we illustrate the process of electron pumping from the localized center to the conduction band where the center relaxes to a metastable state and the subsequent process of electron recapture through a barrier energy $E_{th}$. However, it should be mentioned here that subsequent prolonged and repeated photo-excitation led to stable electron conduction that does not decay even after heating up to 400 °C. The decay of the induced conductivity and its dependence on photon energy, and intensity and illumination time are discussed in detail in the following.

Figure 4 shows the electrical conductivity of the sample as a function of illumination time during exposure to 400 nm (3.1 eV) excitation for a long period of time. The conductivity abruptly increases at the beginning and remains almost constant after exposing the sample to light for 40 minutes. The exposure was continued for 70 hours and then turned off. The decay of conductivity with time after turning the light off after 1-hour excitation is shown in Fig. 5a. For photo-excitation of 3.39 and 3.22 eV, an immediate decline in electron conduction was observed after turning off the light. However, in the case of 3.1 eV excitation, the conductivity gradually decreased and attained the value of dark conductivity after 8 minutes. This indicates the presence of more than one localized state in the band gap and that the 3.1 eV light most likely excites an electron from a center that exhibits a metastable state with longer decay time. Figure 5b (the blue curve) shows the decay of conductivity as a function of time after the 70 hours exposure to 3.1 eV excitation. After decay, the sample was then re-exposed to 3.1 eV photons for 10 minutes. The red curve in Fig. 5b represents the subsequent decay of conductivity. Each of the decay curves in Fig. 5b exhibits two time decay constants, one relatively fast and one slow. The fast decay rate of conductivity was 0.4 $\Omega^{-1}cm^{-1}$/min after 1 h excitation, 0.008 $\Omega^{-1}cm^{-1}$/min after 70 h excitation and 0.004 $\Omega^{-1}$ $cm^{-1}$/min after repeating photoexcitation with higher intensity. The dependence of photoconductivity decay on excitation time is unusual. However these measurements clearly demonstrate the strong dependence of decay rate of conductivity on the energy, intensity and time of photo-excitation. It is interesting that repeated exposure to light significantly impacts the decay of conductivity, leading to a more stable electron conduction.

To further investigate the conditions that cause the permanent transition from insulator to conductor state, an undoped $Ga_2O_3$ sample was exposed to light several times and the conductivity was monitored. Initially the conductivity was $1.08 \times 10^{-8}$ $\Omega^{-1}.cm^{-1}$. When exposed to photo-excitation of 3.1 eV, the conductivity promptly increased by almost two orders of magnitude but retained nearly the same initial value after the light was turned off. However, by repeating photo-excitation and after prolonged exposure to light, the conductivity was increased by 9 orders of magnitude and was held after turning off the light without decay, indicating a complete conversion from the insulator to conductor state. Annealing the sample at 400 °C for 1 hour in dark did not remove or decrease the conductivity. Annealing at a much higher temperature of 800 °C for 2 hours in $O_2$ flow was necessary to revert the sample to an insulator with a conductivity of $7.69 \times 10^{-7}$ $\Omega^{-1}.cm^{-1}$. However, this annealing also completely eliminated the photoconductivity feature of the sample. Figure 6 summarizes these results.

Defects are thought to provide localized states in the bandgap and lead to persistent photoconductivity. The unusual permanent conversion from insulator to conductor observed here and the ability to eliminate this effect by annealing in $O_2$ at high temperatures confirm the significant role of defects. To investigate the presence and nature of defects in $Ga_2O_3$ sample, positron annihilation spectroscopy (PAS) which is a well-established technique to probe vacancy type defects[28] was carried out. Positron annihilation lifetime measurements (PALS) were performed using gamma-induced positron spectroscopy (GIPS) at the ELBE (Electron Linac with high Brilliance and low Emittance) facility, at the Helmholtz-Zentrum Dresden-

Rossendorf (HZDR) in Dresden, Germany. GIPS is an advanced PAS technique that can generate a positron decay curve free from background or source contributions[29]. It uses high-energy γ-rays to generate positrons directly inside the sample by pair production. The main advantage is that it completely eliminates unwanted contributions from positron annihilation in either the source or cladding materials and thus results in accurate measurements of positron lifetimes[30]. PALS measurements on the undoped $Ga_2O_3$ samples used in this work reveal a lifetime of 187± 1 ps. This relatively short lifetime cannot be associated with Ga vacancies which strongly trap positrons, leading to a much longer lifetime. Compared to the reported bulk positron lifetime value of 175 ps in $Ga_2O_3$[31, 32], the 187 ps measured here is a modest increase above the bulk positron lifetime, a result that often indicates the presence of oxygen vacancies in oxides.

Based on PAS measurements and oxygen-annealing experiments, we interpret the persistent photoconductivity in these undoped $Ga_2O_3$ crystals to the presence of large concentrations of oxygen vacancies. We explain the results as follows: an oxygen vacancy $V_O$ in its neutral charge state forms a localized occupied deep state in the band gap and does not lead to conductivity. By exposing the sample to sub-band gap light, electrons are pumped to the conduction band through two excitation steps producing a $V_O^{2+}$ state, which may provide shallow states. In fact, the temperature dependence of the induced conductivity and the electron density presented in Fig. 7 shows a freeze out region for the electrons indicating that these new states are still within the bandgap.

To understand the reason behind the permanent conversion from insulator to conductor and reveal the mechanism that prevented the electrons from returning to their center after turning off light, the change in the structural properties of β-$Ga_2O_3$ has been examined by first-principles electronic structure calculations. There are two different types of Ga sites present in the β-Ga$_2$O$_3$ crystal structure. The first is Ga coordinated by four oxygen (denoted as Ga$_1$) while the second is Ga coordinated with six oxygen (denoted as Ga$_2$ as shown in Fig. 8) The structure also possesses three inequivalent oxygen sites, two 3-fold coordinated O(I) and O(II) sites, and one 4-fold coordinated O(III) site. Out of these three different sites, the O(II) site has the lowest formation energy for a neutral oxygen vacancy[35]. Therefore, in this study we have focused on O(II) type vacancy to evaluate the structural distortion due to the net charge on the defect. It is observed that in the presence of the neutral ($V_O^X$) and charged ($V_O^{\ddot{}}$) oxygen vacancy, the Ga-triangle (formed by the vacancy surrounded by three Ga atoms, denoted by the black dotted line in Fig. 8) contracts and expands respectively as the net charge on the vacancy is changed from neutral to positive. The average Ga-Ga bond length is 3.129 Å and 4.311 Å, respectively, for the neutral and charged cases. (For comparison, in the pristine (non-defective) structure, the average Ga-Ga bond length is 3.306 Å.) More specifically, in the case of $V_O^X$, two Ga$_1$ type atoms relax towards the vacancy site while the Ga$_2$ type atom relaxes away from the vacancy. In contrast, in the $V_O^{\ddot{}}$ case, both the Ga$_1$ and Ga$_2$ type atoms relax away from the vacancy. The variation of the distance of Ga atoms from the vacancy with respect to the pristine structure is shown in Fig. 8. Therefore, inward and outward relaxation of the Ga-triangle decreases (for $V_O^X$) and increases (for $V_O^{\ddot{}}$) the average Ga-Ga bond length with respect to the pristine case.

To scrutinize the structural distortion around the vacancy, we analyzed the electron localization function (ELF) in the three systems. The ELF gives a direct space representation of the electron distribution, which is useful for examining bonding features. The local value of the ELF at a given position can be interpreted as the probability of finding an electron at that locality given the existence of neighboring electrons. The value of the ELF ranges from 0 to 1. ELF values close to one suggest a region of space with high probability of finding electron localization, whereas a value of zero corresponds to a region where either the electron is fully delocalized or does not reside. Finally, an ELF value close to one-half implies that the region exhibits electron gas-like behavior. Two-dimensional ELF contour plots of pristine $Ga_2O_3$ and $Ga_2O_3$ containing a neutral vacancy and a charged vacancy are shown in Fig. 9. The electron localization on Ga is lower than on O, but there is strong overlap of electron localization. Redistribution of the electron localization probability is depicted for the case of the neutral O vacancy in Fig. 9b, where a strong probability of finding an electron is observed between two Ga atoms near the vacancy site. This reduces the $Ga_1$-$Ga_2$ bond length and pushes $Ga_1$ towards the defect site. The $Ga_2$ atom also interacts slightly with this localized charged state, but it has a much stronger interaction with the neighboring O, which pull $Ga_2$ away from the defect site. In the charged vacancy case (Fig. 9c), an ELF quite similar to the pristine case is observed except around the defect site. Moreover, the ELF does not exhibit the localized charge state between two $Ga_1$ atoms. Rather, the $Ga_1$ atoms are displaced from their original positions and move away from the defect site to form a new $Ga_1$-O bond. However, the ELF of the $Ga_2$ atom is similar to the neutral vacancy case. This results in the massive expansion of the Ga triangle and corresponding increase in $Ga_1$-$Ga_1$, $Ga_1$-$Ga_2$ bond lengths.

The total and partial electronic density of states (DOS) are plotted in Fig. 10 to provide insight about the impact of the structural relaxation of neutral and charged vacancy on the electronic states within the material. The DOS of the pristine structure is also shown for comparison. It is observed in Fig. 10 that the presence of the neutral vacancy introduces an occupied localized state just above the valence band due to the internal inward relaxation shown in Fig. 8b. On the other hand, in the case of the charged oxygen vacancy, the outward relaxation of the Ga-triangle shifts the defect states to higher energy, closer to the conduction band minimum region (in the DOS in Fig. 10, these overlap with the conduction band in a way that makes it difficult to discern by eye). Those states are now unoccupied.

These calculations reveal that changing the charge state of the vacancy leads to a strong structural relaxation and, as would be expected, a change in the defect electronic states in the bandgap. They confirm the experimental scenario that the vacancy structure changes when it is excited. Once the vacancy charge state changes, not only is the defect state emptied, but the state shifts significantly towards the conduction band, leading to a situation in which there is no energetic driving force for the excited electrons to re-encounter the vacancy.

**Conclusion**

We have revealed a mechanism for insulator conductor transition through the redistribution of electron localization in the lattice induced by altering the charge states of defect centers and the subsequent drastic lattice distortion and large shift in the density of states. The measurements demonstrated that sub-band gap light illumination of undoped $Ga_2O_3$ for limited time leads to a permanent transition from a highly insulating state to a conductive state that cannot not reversed, an unusual surprising phenomenon with vast implications on both the properties and potential applications of the material. We propose that such mechanism may take place in other wide band gap oxides strongly impacting their properties and applications. We also propose that the dependence of the decay of conductivity on the photo-excitation time and intensity revealed in this work may open up a new frontier to tuning material properties and developing devices that can be controlled by light. By limiting the excitation time or decreasing the photon intensity, conductivity can be generated and erased providing opportuning for optical memory uses. Further, long time excitation can be used to develop n-type semiconductor for electronics.

# Methods

Czochralski (CZ) grown un-doped, Fe-doped, and Mg-doped bulk single crystals of $Ga_2O_3$ were obtained from Synopsis Inc. The as-grown crystals were about 1cm in diameter and were sliced into pieces of 1 mm thickness. The electrical transport properties of the samples were measured using a MMR Hall effect measurement system. Before the measurement, samples were properly cleaned, and indium contacts were mounted on the surface of each samples. Light emitting diodes (LEDs) of various wavelength (365, 385, 400, 460, 650 and 850nm) were used to provide photo-excitation of 3.39, 3.22, 3.1, 2.69, 1.9 and 1.45 eV, respectively, and the photo-Hall measurements were carried out at room temperature. The intensity of photo-excitation was varied by changing the current passing through the LEDs. For the photo-Hall measurements, the Hall-effect chamber is customized with a transparent window for the illumination of the sample and a Joule Thompson refrigerator is used to keep the sample temperature constant, overcoming the heating effect caused by light illumination. The refrigerator operates by running high pressure nitrogen gas through thin pipes. Because of the light induced heat, this setup is crucial for photoconductivity experiments to investigate the change in carrier concentration and conductivity due solely to photoexcitation, without the influence of thermal contributions. Temperature dependent Hall effect measurements were performed on the permanent state conductive sample from 10-300K using a cryostat with He compressor.

## Density Functional Theory Calculations

The structural properties of β-$Ga_2O_3$ have been investigated by the density functional theory (DFT) formalism as implemented in VASP (Vienna Ab initio Simulation Package)[33] . The core electron behaviour and the interaction between the valence electrons and the ion are described by the projector augmented wave method

(PAW)[34]. The Perdew–Burke–Ernzerhof (PBE) form of the generalized gradient approximation (GGA) has been employed as the exchange-correlation functional to obtain the optimized ground state structure[35]. The Brillouin zone has been sampled using 3x3x3 and 7x7x7 meshes of Monkhorst–Pack $k$-points for optimization and electronic structure calculations, respectively. The valence electrons are described by a plane waves basis set with a converged energy cut-off of 520 eV. A supercell of 160 atoms (32 formula unit) is considered in this calculation. The structure has been optimized until the calculated Hellmann–Feynman forces are smaller than 0.0001 eV Å$^{-1}$.

## Author Contributions

FAS designed and interpreted the experiments, DR, SA, MI, PS, PD, NA carried the annealing and conductivity experiments, MB, MOL, AW carried PALS experiments, AB and BPU performed DFT calculations and analysis, FAS and BPU wrote the manuscript.

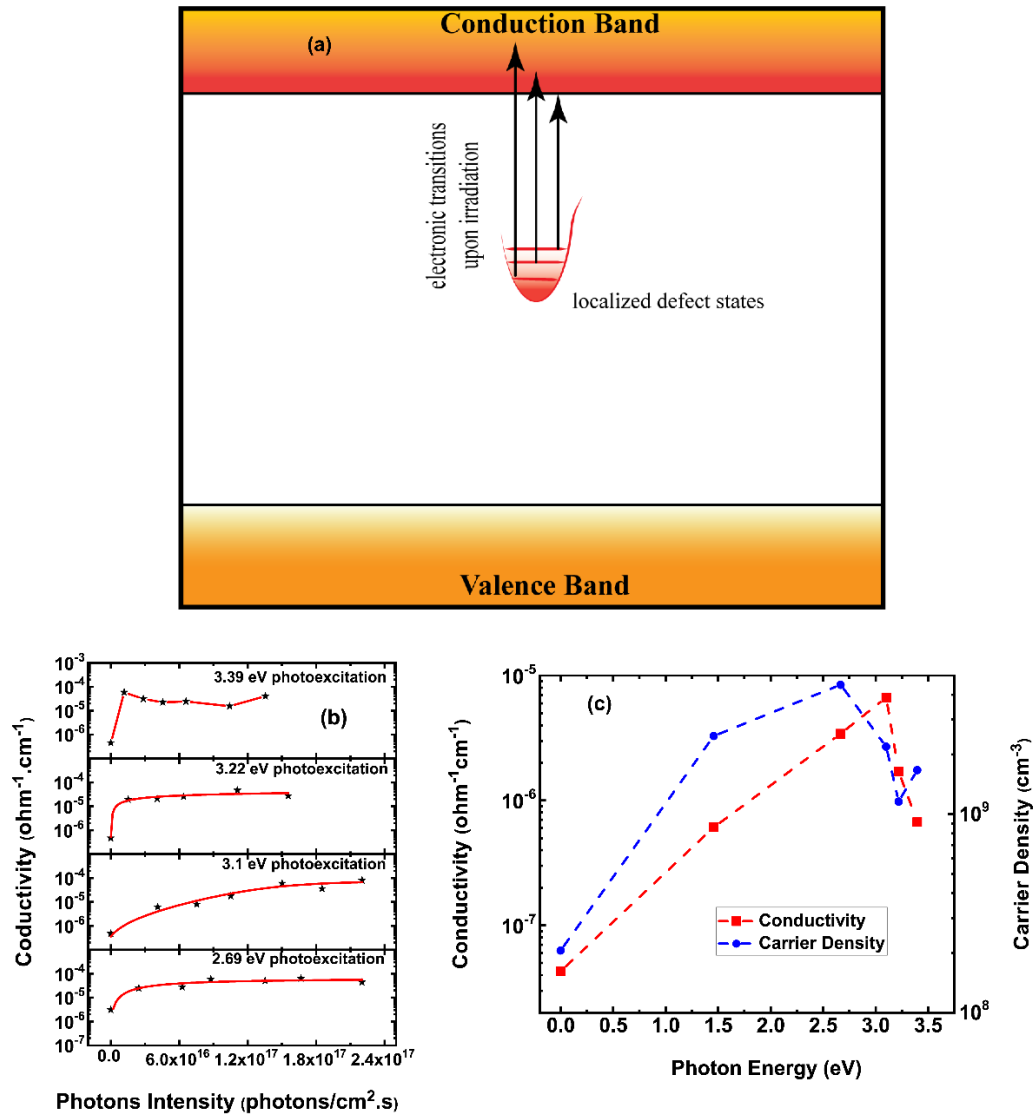

Figure-1: Photoconductivity in undoped β-Ga$_2$O$_3$ bulk crystals. (a) A schematic showing light-induced transition of electrons from localized states within the gap to the conduction band. (b) Dependence of photo-conductivity on photo-excitation energy and intensity at room temperature. The initial dark conductivity before illuminating with 2.69 eV photons was on the order of 10$^{-6}$ ohm$^{-1}$.cm$^{-1}$, which is one order of magnitude higher than the initial value in other measurements. This increase in dark conductivity was due to the exposure of the sample to room light. (c) Dependence of photo-conductivity and photo-induced charge carrier density on photon energy at room temperature.

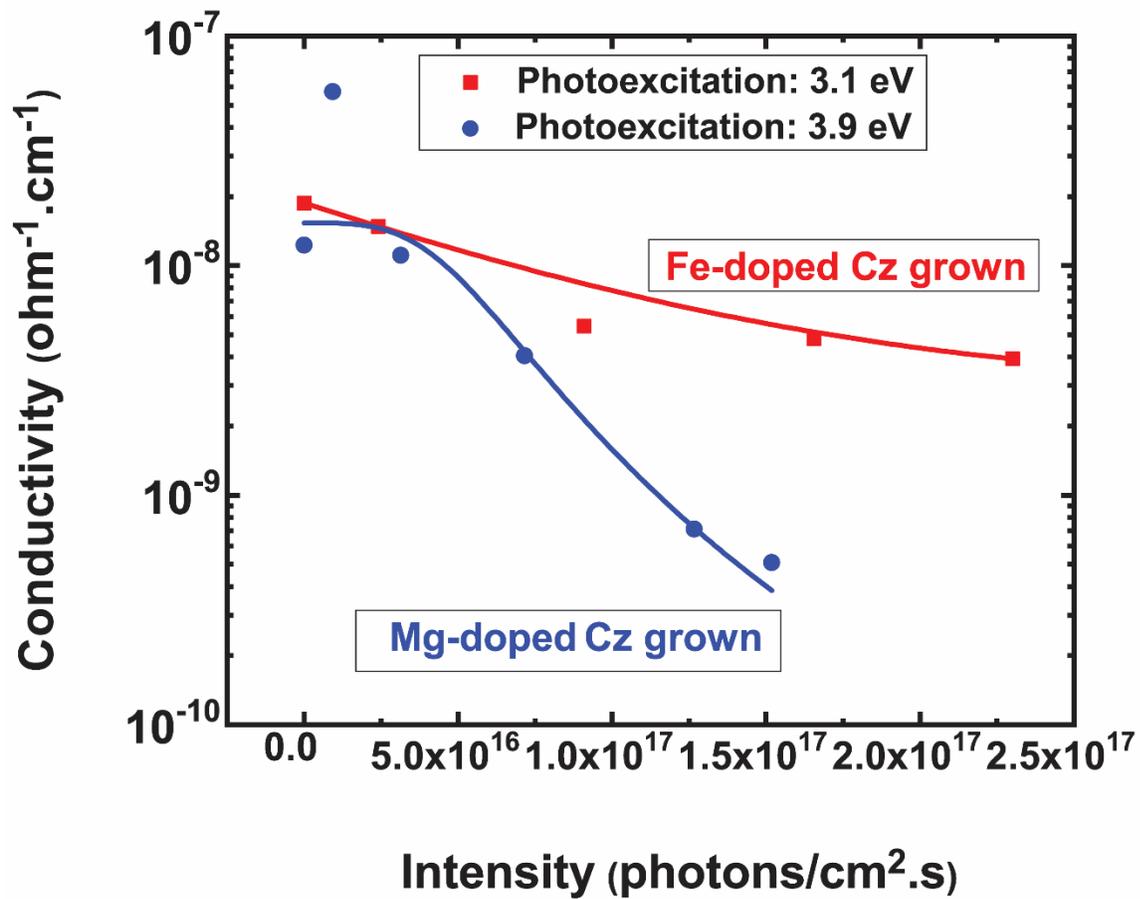

Figure-2: Change in conductivity as a function of photon intensity in doped β-Ga$_2$O$_3$ bulk crystals: Fe-doped Ga$_2$O$_3$ and Mg- doped Ga$_2$O$_3$. Contrary to undoped crystals, the conductivity here decreased by exposing the samples to light, illustrating that impurities such as Fe and Mg do not play a role in the induced photoconductivity of Ga$_2$O$_3$.

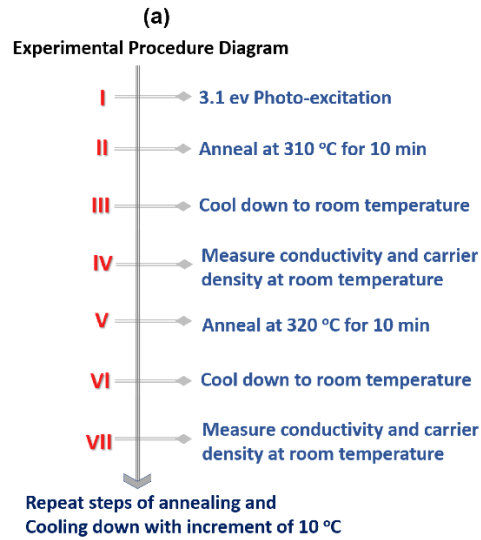
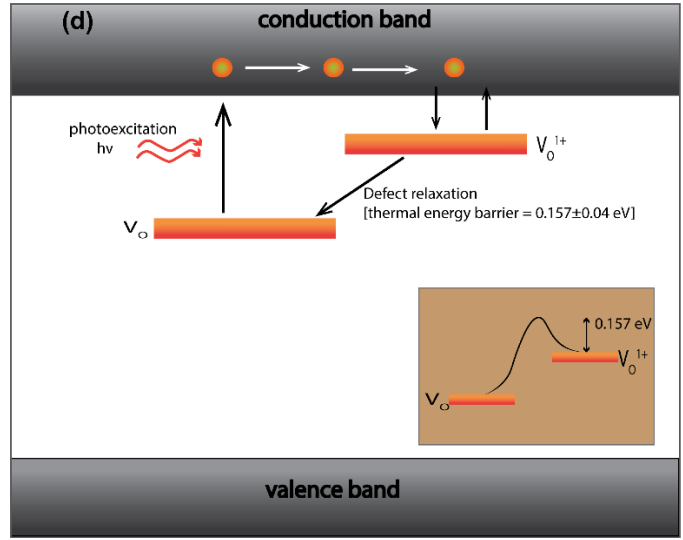
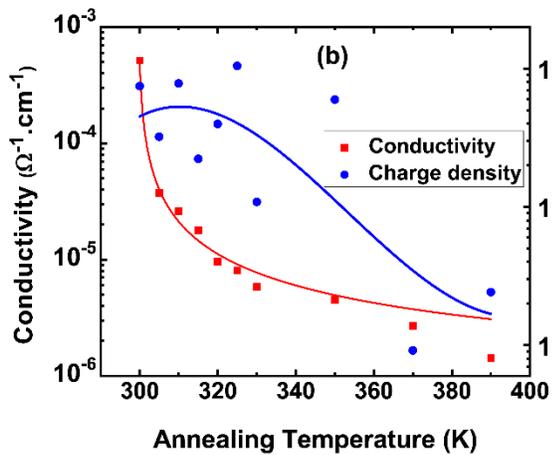
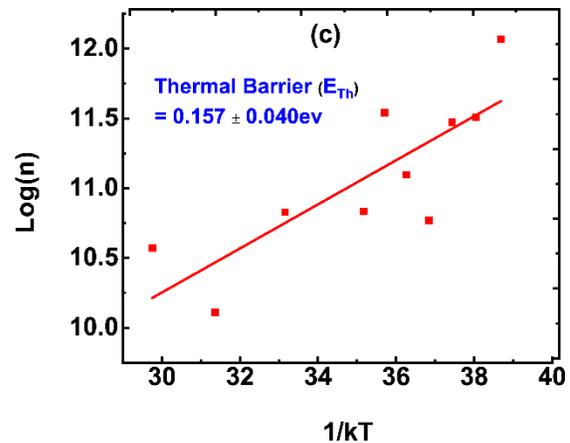

Figure-3: Decrease of conductivity and charge carrier density after photo-excitation with annealing temperature and calculation of the thermal barrier energy for electron recapture by defect. (a) A diagram showing the experimental work flow. (b) Conductivity and Carrier density versus annealing temperature. (c) Log of carrier density vs $1/kT$, where k is the Boltzmann constant and T is temperature in Kelvin. (d) A schematic showing the excitation of an electron from the defect state to the conduction band and the thermal barrier energy for electron recapture.

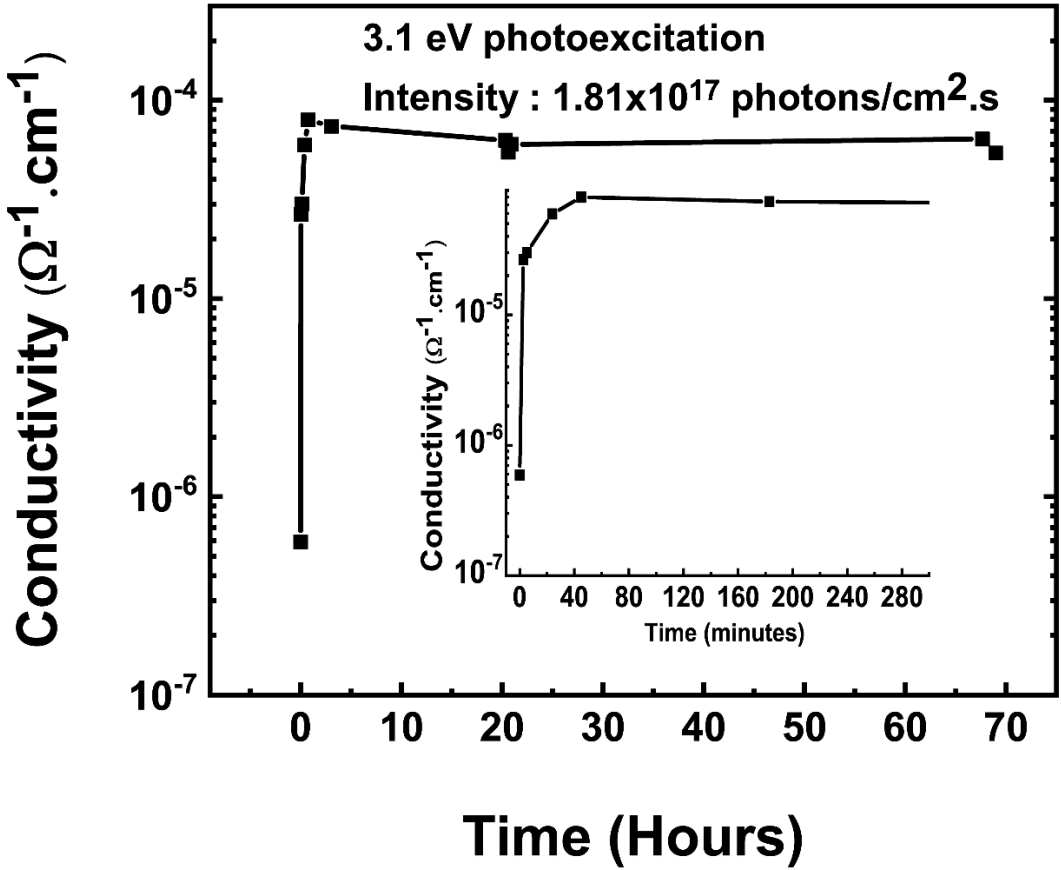

Figure-4: Electrical conductivity as a function of illumination time (in-situ measurements with light on). The inset highlights the increase in conductivity with illumination in the first several hours showing saturation around 40 minutes.

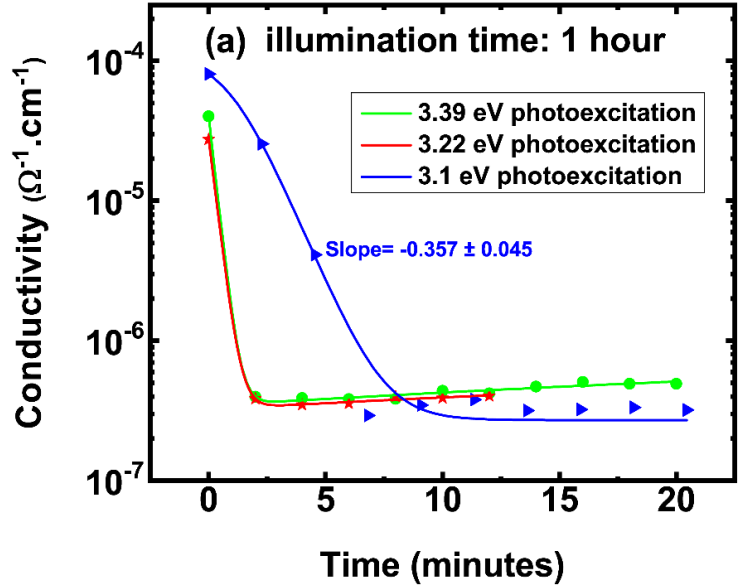
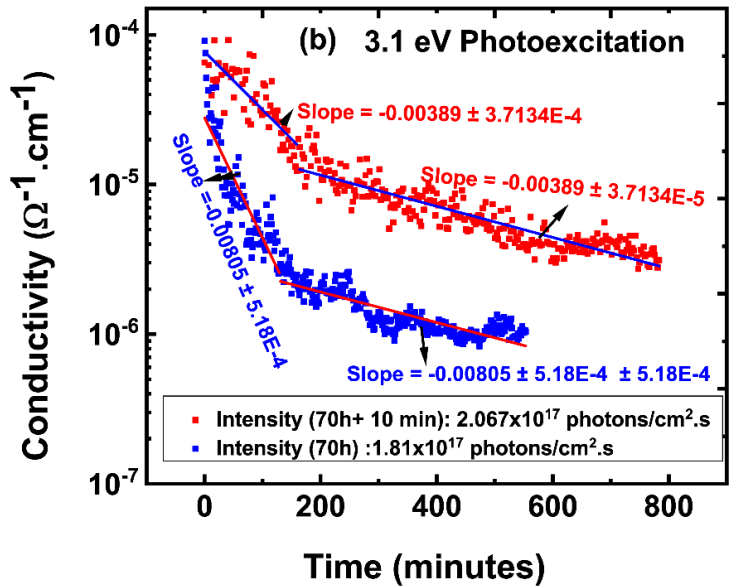

Figure-5: Decay of photo-conductivity as a function of time after turning off photoexcitation. (a) After exposing the sample to light with different photon energies for one hour. (b) After exposing the sample to 3.1 eV photoexcitation. The blue curve represents the decay in conductivity after illuminating the sample for 70 hours. After the initial decay in conductivity, the sample was exposed again to higher photon intensity for about 10 minutes (red curve). These results illustrate the dependence of the conductivity decay rate on the energy, intensity and time of photoexcitation as well as on the repeated exposure to light.

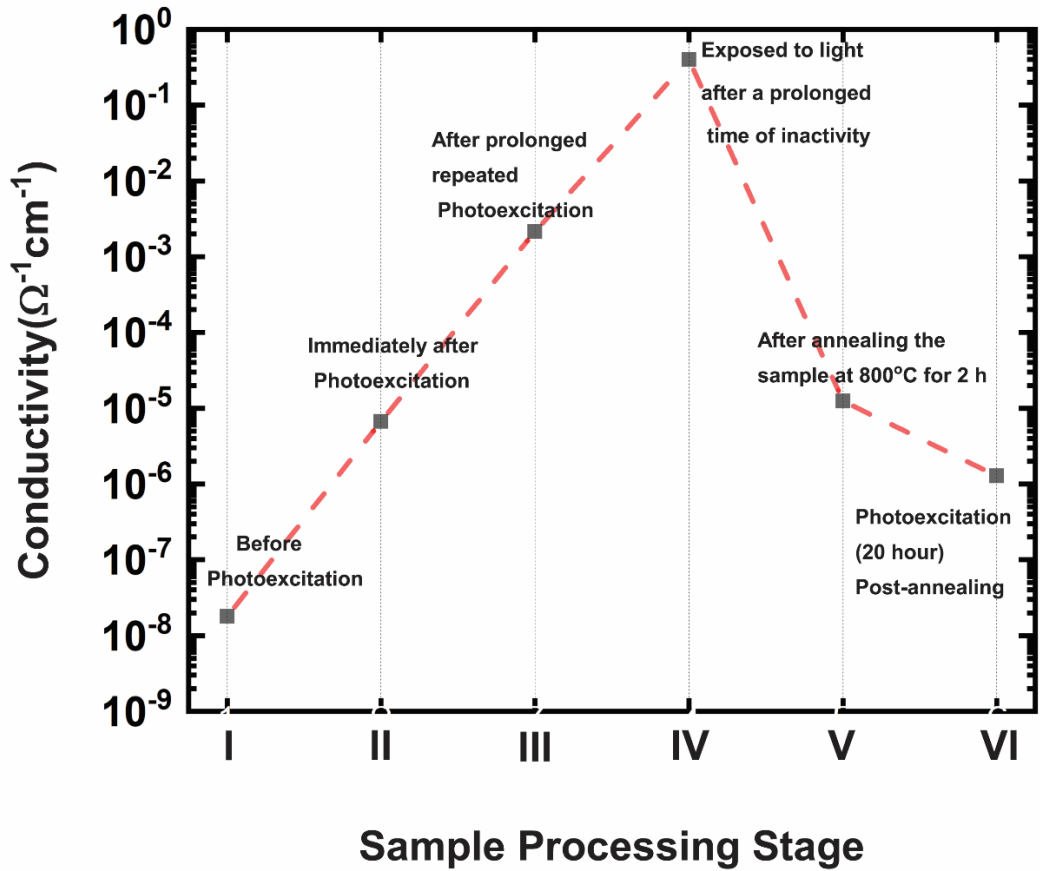

Figure-6: Change in electron conduction in undoped β-Ga$_2$O$_3$ after exposing to photo-excitation and after annealing. After repeated photo-excitation at 3.1 eV, the conductivity increased 9 orders of magnitude from 10$^{-8}$ to almost 1 Ω$^{-1}$ cm$^{-1}$ and retained this conductivity without decay, indicating a transition from insulator state to conductor state. Then, the conductivity was suppressed to about 10$^{-6}$ Ω$^{-1}$ cm$^{-1}$ by annealing the sample at 800 °C in O$_2$ for 2 hours. After annealing, prolonged re-photoexcitation for 20 hours did not increase the conductivity, on the contrary it further decreased it illustrating that the sample lost all its photo-conductivity feature by high temperature anneal in O$_2$.

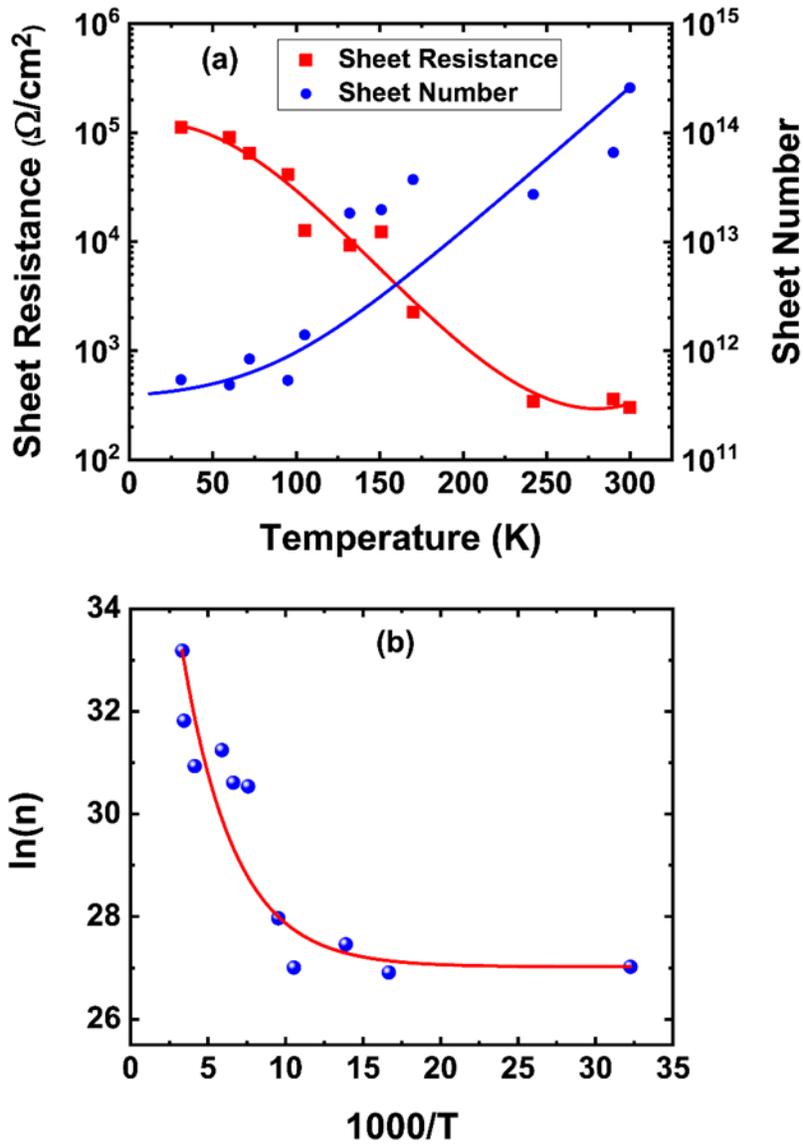

Figure-7: Electronic conduction-temperature characteristics of the permanent state conductivity in $Ga_2O_3$: (a) sheet resistance and sheet number as a function of temperature and (b) Ln of sheet number as a function of inverse temperature. The measurements reveal a freeze out region for charge carrier indicating that the induced new states are shallow states within the band gap and not in the conduction band.

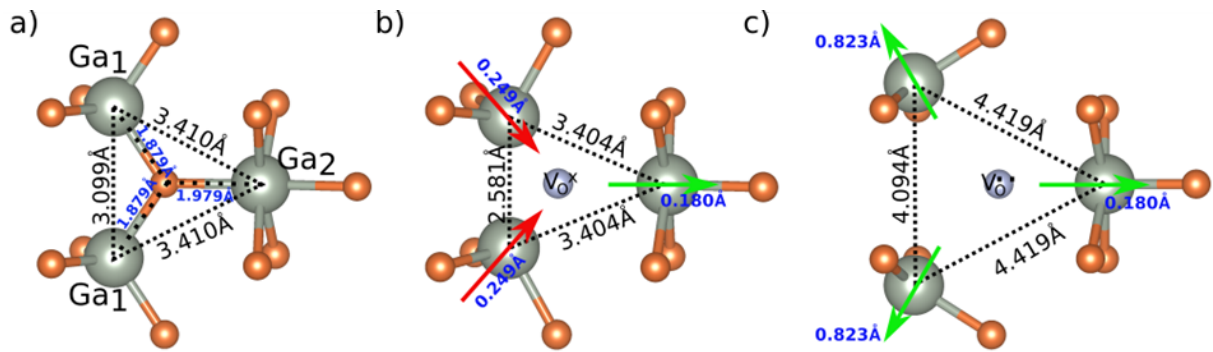

Figure-6: Internal structural arrangements in β-Ga$_2$O$_3$: (a) pristine structure, (b) with neutral oxygen vacancy, and (c) with doubly charged oxygen vacancy. Variation in the distance of Ga atoms from the vacancy is also shown. Red and green arrows with indicated values show how much Ga atoms move towards or away from the vacancy, respectively. Grey and orange balls represent Ga and O atoms, respectively.

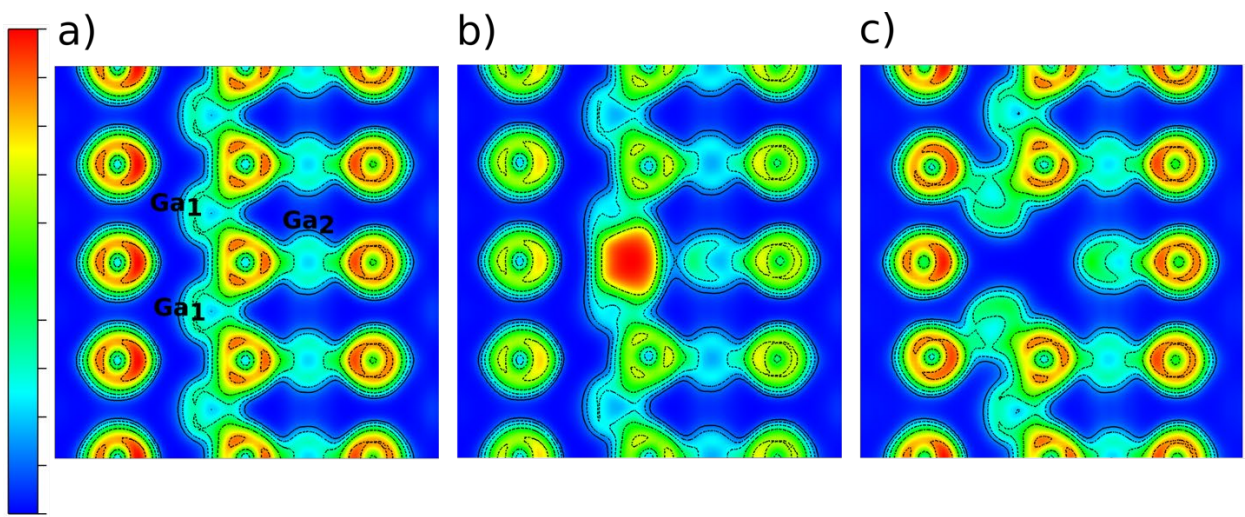

Figure-7: Two-dimensional contour plots of the electron localization function (ELF) for (a) pristine Ga$_2$O$_3$, b) Ga$_2$O$_3$ containing a neutral oxygen vacancy, and c) Ga$_2$O$_3$ containing a 2+ charged oxygen vacancy. Blue, green and red correspond to ELF values of 0, 0.5 and 1, respectively.

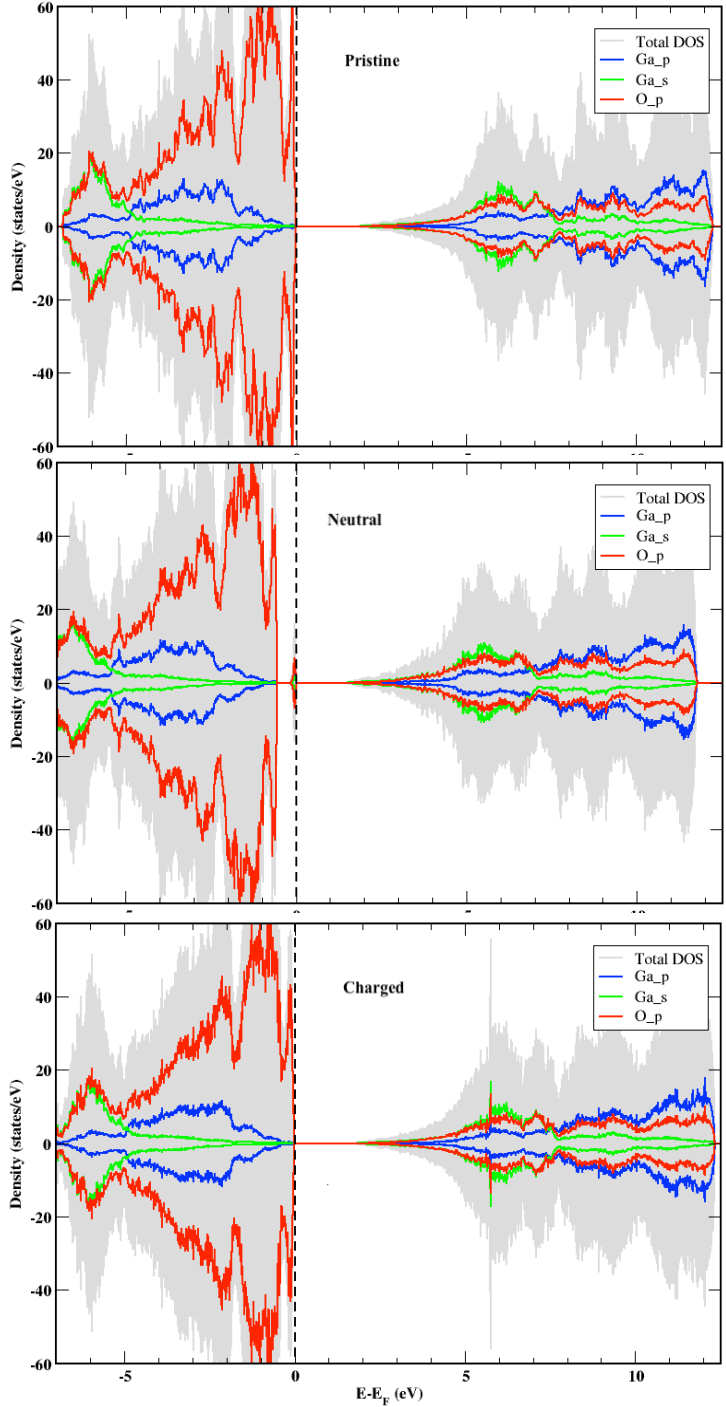

Figure-8: Total (gray) and partial (lines) electronic density of states (DOS) of (a) pristine $Ga_2O_3$ and $Ga_2O_3$ with (b) a neutral and (c) charged oxygen vacancy. The s and p partial DOS for Ga and the p partial DOS for oxygen are represented by the green, blue and red lines, respectively. The zero of energy is referenced against the highest occupied state in the system, referred to here as the Fermi energy $E_F$.